# Low temperature magnetic properties of Nd$_2$Ru$_2$O$_7$

S. T. Ku[1], D. Kumar[1], M. R. Lees[2], W-T. Lee[3], R. Aldus[3], A. Studer[3], P. Imperia[3], S. Asai[4], T. Masuda[4], S. W. Chen[5], J. M. Chen[5†], L. J. Chang[1,6*]

[1]Department of Physics, National Cheng Kung University, Tainan 70101, Taiwan

[2]Department of Physics, University of Warwick, Coventry CV4 7AL, United Kingdom

[3]Australian Nuclear Science and Technology Organization, Locked Bag 2001, Kirrawee DC, NSW 2232, Australia

[4]Institute for Solid State Physics, University of Tokyo, Kashiwa, Chiba 277-8581, Japan

[5]National Synchrotron Radiation Research Center, Hsinchu 30076, Taiwan

[6]Advanced Science Research Center, Japan Atomic Energy Agency (JAEA), Tokai, Naka, Ibaraki 319-1195, Japan

[*]ljchang@mail.ncku.edu.tw, [†] jmchen@nsrrc.org.tw

**Abstract**

We present magnetic susceptibility, heat capacity, and neutron diffraction measurements of polycrystalline Nd$_2$Ru$_2$O$_7$ down to 0.4 K. Three anomalies in the magnetic susceptibility measurements at 146, 21 and 1.8 K are associated with an antiferromagnetic ordering of the Ru$^{4+}$ moments, a weak ferromagnetic signal attributed to a canting of the Ru$^{4+}$ and Nd$^{3+}$ moments, and a long-range-ordering of the Nd$^{3+}$ moments, respectively. The long-range order of the Nd$^{3+}$ moments was observed in all the measurements, indicating that the ground state of the compound is not a spin glass. The magnetic entropy of $R$ln2 accumulated up to 5 K, suggests the Nd$^{3+}$ has a doublet ground state. Lattice distortions accompany the transitions, as revealed by neutron diffraction measurements, and in agreement with earlier synchrotron x-ray studies. The magnetic moment of the Nd$^{3+}$ ion at 0.4 K is estimated to be 1.54(2)$\mu_B$ and the magnetic structure is all-in all-out as determined by our neutron diffraction measurements.

## 1. Introduction

Frustrated pyrochlore compounds with the general formula $A_2B_2O_7$ [1-7], where $A$ is a trivalent rare-earth ion and $B$ is a tetravalent transition-metal ion [1], with space group $Fd$-$3m$, are made up of interpenetrating $A_2$O′ and $B_2$O$_6$ sublattices composed of corner-sharing tetrahedra. Pyrochlore oxides have been extensively studied in the past few decades and various novel magnetic behaviours have been discovered including spin glass [2], spin ice [3,4], and spin liquid [5]. These different magnetic states arise from the geometrical frustration of the pyrochlore lattice, which can lead to a macroscopically degenerate magnetic ground state rather than a conventional long-range ordered state. To date, many of the experimental results have been obtained on heavy rare-earth titanate pyrochlores possessing large *4f* magnetic moments. These include $Dy_2Ti_2O_7$ [6] and $Ho_2Ti_2O_7$ [4] identified as spin ice, and $Tb_2Ti_2O_7$ [7] known as spin liquid. In spin ice materials, the ferromagnetic interactions between the magnetic moments sitting on the corners of the tetrahedra become frustrated because of the crystalline electric field (CEF). These crystal fields require the magnetic moments on the corners of the tetrahedra to align along the local cubic <111> directions, in such a way that the moments can point either in or out of the centre of the each tetrahedron. In spin-ice materials, this "two-in/two-out" spin configuration is analogous to the proton arrangement in water ice [4]. Spin ice has been suggested to host unconventional low temperature magnetic and thermodynamic properties, including the Dirac strings observed in $Dy_2Ti_2O_7$ [8], while the quantum spin-ice candidate $Yb_2Ti_2O_7$ is suggested to exhibit a Higgs transition [9]. In other pyrochlores, where the exchange interactions are antiferromagnetic and stronger than any dipolar interactions, long-range antiferromagnetic ordering is observed [10].

Recently, interest has moved to the pyrochlore compounds with smaller moments, in which different quantum effects such as novel low-energy spin dynamics and planar components to the spin symmetry have been observed [11]. Since the $Nd^{3+}$ ion possesses a relatively small magnetic moment, the strength of the dipole-dipole interaction is significantly reduced and the exchange interaction may dominate the system in a similar way to that seen in $Yb_2Ti_2O_7$ and hence, may result in long-range ordering [9]. Ruthenium pyrochlores have attracted a lot of attention over the last few years [12, 13]. For ruthenium pyrochlores, both the *A* (16*d*) and *B* (16*c*) sites are occupied by magnetic ions. The perturbation driven by the moment at the 16*c* site may cause a symmetry breaking and structural distortions [14]. Early studies on the ruthenium pyrochlores revealed that there is a λ-type jump in the specific heat in temperature range 75 to 160 K depending on the size of the rare-earth ion [15, 16]. The magnetic moment per Ru obtained from neutron diffraction [17] is reduced compared to the theoretical maximum expected for $4d^4$ low-spin configuration, *i.e.*, $2\mu_B$, indicating the observed Ru-moment is modified by the CEF. In the case of $Nd_2Ir_2O_7$, a metal-insulator transition has been observed near 33 K and an all-in/all-out magnetic structure has been suggested by neutron diffraction and muon spin relaxation measurements [18, 19]. Long-range magnetic ordering below 0.55 K has been confirmed in $Nd_2Hf_2O_7$ by powder neutron diffraction measurements, with an all-in/all-out magnetic structure and an ordered magnetic moment of $0.62\mu_B$ at 0.1 K [20]. $Nd_2Zr_2O_7$ has also been confirmed to show long-range magnetic ordering below 0.4 K with an all-in/all-out magnetic structure [21].

In published reports, the anomalies observed in the heat capacity and magnetic susceptibility measurements at 146 and 21 K in $Nd_2Ru_2O_7$ were associated with the Ru antiferromagnetic order temperature and a glassy-like transition, respectively [15, 16]. In the neodymium pyrochlore, the $Nd^{3+}$ ion site has trigonal symmetry and the crystalline electric field splits the ground state manifold into 5 Kramers doublets leading to a calculated magnetic moment of $3.62\mu_B$ [20]. One of the interesting features of pyrochlores with a Kramers doublet, which have a well-separated ground state and first excited state, is that the ground state properties can be explained by the pseudo-spin $S = ½$. [20]. In particular, the Kramers doublet systems with total angular momenta $J = \frac{9}{2}$ and $\frac{15}{2}$ (for $Nd^{3+}$) are very interesting as it has been observed that under certain conditions these doublets may behave like "dipolar-octupolar" doublets [22], making $Nd_2Ru_2O_7$ an interesting compound for further studies.

Here we have studied the behaviour of $Nd_2Ru_2O_7$ at low temperatures, in order to investigate the magnetic ground state. We report dc magnetic susceptibility $\chi_{dc}$, ac magnetic susceptibility $\chi_{ac}$, and heat capacity $C_p$ measurements on $Nd_2Ru_2O_7$ as a function of temperature and magnetic field. The cubic *Fd-3m* pyrochlore structure is confirmed by powder x-ray diffraction (XRD) measurements. The $Ru^{4+}$ moments order at 146 K, followed by an anomaly at 21 K due to a canting of the $Ru^{4+}$ moments and perhaps a polarization of the $Nd^{3+}$ spins. Magnetization and heat capacity measurements reveal the long-range order of the $Nd^{3+}$ spins at 1.8 K due to $Ru^{4+}$ - $Nd^{3+}$ coupling. Powder neutron diffraction studies have confirmed antiferromagnetic ordering of the Nd moments below 1.8 K, and reveal an all-in/all-out magnetic structure in $Nd_2Ru_2O_7$ with an ordered magnetic moment of $1.54(2)\mu_B$ at 0.4 K. To the best of our knowledge, this is the first report of long-range order of the $Nd^{3+}$ moments in $Nd_2Ru_2O_7$ below 1.8 K.

## 2. Experimental methods

Polycrystalline samples of $Nd_2Ru_2O_7$ were synthesized using stoichiometric quantities of high purity starting materials $Nd_2O_3$ (99.99%) and $RuO_2$ (99.99%) from Alfa Aesar. The powder materials were thoroughly mixed together by grinding with a mortar and pestle for more than an hour. The samples were then pre-sintered at 800 ˚C for 12 hours. The finely ground powders were then pelletized using an isostatic cold press and sintered at 1150 ˚C for a period of four days with several intermediate grindings. We used alumina crucibles and the heat treatments were performed in air.

The crystal structure at the room temperature and phase purity of the samples were verified via powder x-ray diffraction (XRD) using Bruker AXS Gmbh D2 Phaser desktop x-ray diffractometer. DC magnetization measurements were performed using a commercial superconducting quantum interference device (SQUID) magnetometer (Quantum Design MPMS) above 1.8 K, while an iQuantum $^3$He insert was also used for the measurements below 1.8 K. Heat capacity measurements were carried out using a relaxation method in a Quantum Design Physical Property Measurement System (PPMS) equipped with a dilution refrigerator insert. Powder neutron diffraction measurements were performed using a $^3$He fridge on the WOMBAT high-intensity diffractometer at ANSTO, Australia using a neutron wavelength of 2.41 Å. The sample was loaded in a Cu can. The $^3$He fridge was used to cool down the sample to 0.4 K and the diffraction patterns were taken at 0.4 and 8 K. The neutron diffraction measurements at high temperatures ($T > 8$ K) were performed with a CF-8 cryofurnace on the WOMBAT diffractometer.

## 3. Results and discussion

### 3.1 Structural studies

Fig. 1(a) shows the results of a Rietveld refinement of the powder XRD pattern of $Nd_2Ru_2O_7$ at room temperature. The data confirm the face-centred cubic (space group *Fd-3m*) pyrochlore structure of $Nd_2Ru_2O_7$ with a lattice parameter of 10.3544(5) Å. All of the observed peaks were indexed successfully, which indicates the sample is single phase to within the detection limit of this technique. Table 1 shows the lattice parameter and crystallographic parameters obtained from the structural Rietveld refinement of the XRD data of $Nd_2Ru_2O_7$. Fig. 1(b) shows the double pyrochlore structure of $Nd_2Ru_2O_7$, where both the *A* ($Nd^{3+}$) and *B* ($Ru^{4+}$) ions are magnetic, formed by two types of corner sharing tetrahedra where one type of tetrahedron has Nd atoms at the corners (shown by the large grey spheres) and the other tetrahedra have Ru atoms at their corners (shown by the small white spheres). The centre of each tetrahedron is occupied by an oxygen atom. The $Nd^{3+}$ ions occupy 16d (½,½, 2) sites and $Ru^{4+}$ ions occupy 16c (0, 0, 0) sites, while the $O^{2-}$ ions occupy two types of anionic sites 8*b* ($\frac{3}{8}, \frac{3}{8}, \frac{3}{8}$) and 48*f* ($x, \frac{1}{8}, \frac{1}{8}$). In the pyrochlore structure, six O1 (denoted by 48*f*) anions are equivalent, while one O2 (denoted by 8*b*) occupies a distinct position in the structure. Therefore, the formula unit of the pyrochlore is represented as $Nd_2Ru_2O(1)_6O(2)$. The Nd and Ru atoms sitting at the corners of the tetrahedra form a three-dimensional network, which leads to the double pyrochlore structure. In this structure, Nd atoms are surrounded by 8 O-atoms, while each Ru atom is surrounded by 6 O-atoms. We have previously published structural studies of $Nd_2Ru_2O_7$ which combine high resolution powder X-ray diffraction (HR-XRD) and extended X-ray absorption fine structure (EXAFS) techniques [14].

The structural variations at the low temperature are reported and a structural instability is observed. As the temperature is decreased below 150 K, anomalies are observed in the lattice parameters and Ru-O, Ru-O-Ru bond distances in $Nd_2Ru_2O_7$ [14].

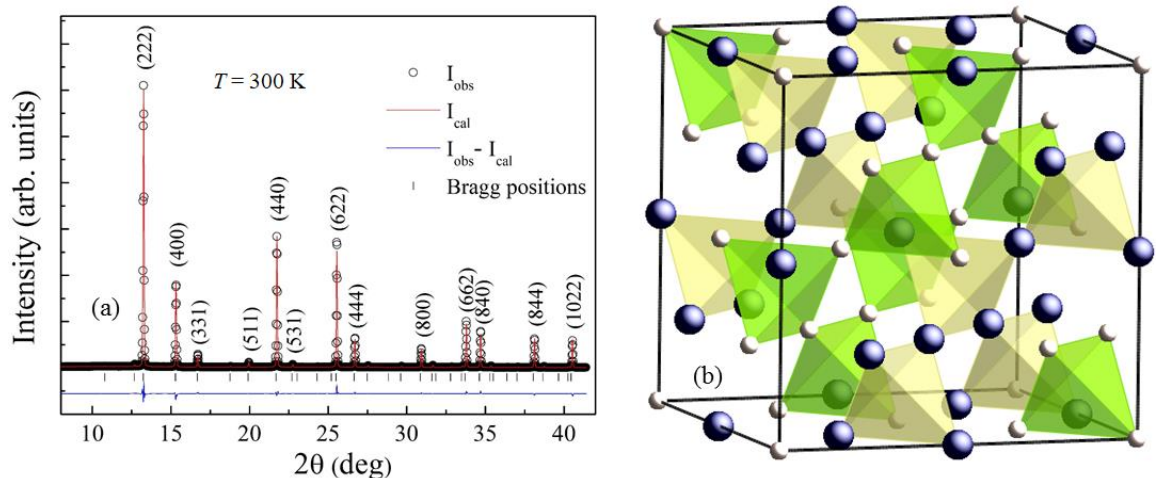

FIG. 1. (a) Rietveld refinement results of the powder x-ray diffraction pattern of $Nd_2Ru_2O_7$ at room temperature. The open circles show the observed intensity data, the red line indicates the theoretically calculated intensity, and the blue line represents the difference between the observed intensity and calculated intensity. The vertical bars mark the Bragg-peak positions. The wavelength of x-rays used is 1.54 Å (b) An illustration of the double pyrochlore structure formed by two types of tetrahedra where one type of tetrahedron has Nd atoms at its corners (shown by the large grey spheres) and the other tetrahedra have Ru atoms at the corners (shown by the small white spheres).

| Lattice parameter | |
| --- | --- |
| $a$ | 10.3544(5) Å |
| Atomic positions | |
| Nd | 0.5000 |
| Ru | 0.0000 |
| O1 | 0.3309 (5) |
| O2 | 0.3750 |
| Refinement quality | |
| $\chi^2$ | 1.12 |
| $R_p$ (%) | 3.99 |
| $R_{wp}$ (%) | 5.70 |

Table 1: Lattice parameter and atomic positions obtained from the Rietveld refinement of the powder XRD data of $Nd_2Ru_2O_7$ at room temperature.

### 3.2 Magnetization measurements

We have studied the magnetic properties of polycrystalline samples of $Nd_2Ru_2O_7$ down to 0.4 K. Fig. 2(a) shows the temperature dependence of the dc magnetic susceptibility, for both field-cooled (FC) and zero-field-cooled (ZFC) data, for $Nd_2Ru_2O_7$ in the temperature range $0.4 \leq T \leq 190$ K. The susceptibility measurement down to 1.8 K were performed with an MPMS squid magnetometer and the measurements below 1.8 K were performed separately using an MPMS magnetometer with a $^3$He insert, as shown by the red zero-field-cooled warming (ZFCW) and blue field-cooled warming (FCW) curves (see inset of Fig. 2a). Various anomalies are observed in the dc susceptibility data.

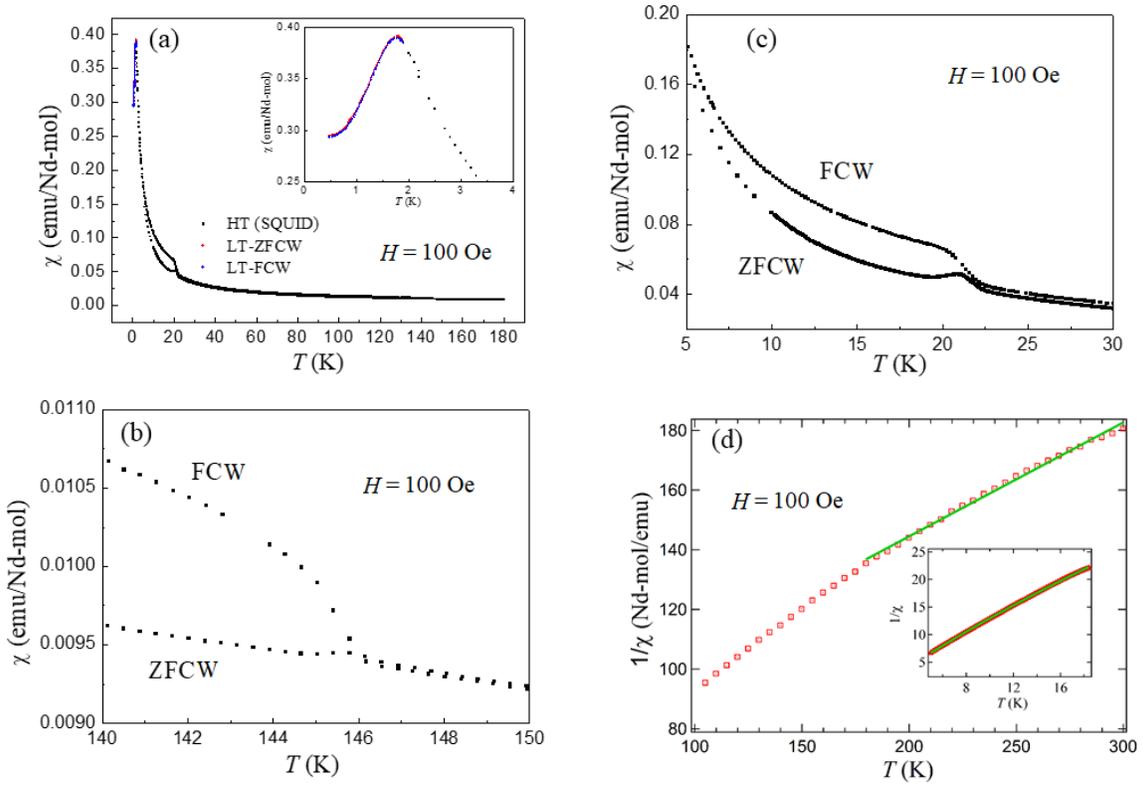

FIG. 2. (a) Temperature variation of the dc magnetic susceptibility of $Nd_2Ru_2O_7$ at $H = 100$ Oe in the temperature range 0.4 to 190 K. Three featuress are observed at 1.8, 21, and 146 K. The inset shows the peak at 1.8 K in more detail. The high temperature (HT) data is taken with an MPMS SQUID magnetometer and the low temperature (LT) data is taken in the same system with a $^3$He insert.. ZFCW and FCW indicate the zero-field-cooled warming and field-cooled warming data respectively. (b) Ordering of the $Ru^{4+}$ $4d$ moments leads to a cusp-like anomaly in the ZFCW magnetic susceptibility and a splitting of ZFCW and FCW curves around 146 K. (c) Temperature variation of the dc magnetic susceptibility of $Nd_2Ru_2O_7$ showing the anomaly at 21 K due to weak ferromagnetic ordering of the $Nd^{3+}$ and $Ru^{4+}$ ions. (d) Temperature variation of inverse susceptibility in the temperature range $100 \leq T \leq 300$ K. The solid line shows a fit to the Curie-Weiss law. The inset shows the inverse susceptibility in the temperature range $5 \leq T \leq 18$ K, where the solid line shows the fit to a modified Curie-Weiss law i.e., $\chi(T) = \chi_0 + C/(T - \theta_{CW})$.

The first anomaly observed around 146 K is due to the antiferromagnetic ordering of the $Ru^{4+}$ spins, where we observe a small cusp-like feature in ZFCW magnetic susceptibility and splitting of ZFCW - FCW curves around 146 K (as shown in Fig. 2b). A corresponding peak is also observed in the specific heat data confirming the $Ru^{4+}$ ordering.

As we decrease the temperature further, we again observe a further splitting of ZFCW and FCW susceptibility curves around 21 K (shown in Fig. 2c). Contradictory conclusions are drawn by reports published in Refs. [16] and [23] regarding the origin of the anomaly at 21 K and the divergence of the ZFCW-FCW curves below 21 K in $Nd_2Ru_2O_7$, which we will discuss later.

As we further cooling, we observe a peak in $\chi(T)$ near 1.8 K (as shown in the inset of Fig. 2a). A corresponding peak is also observed in the specific heat data. This peak is associated with the long-range antiferromagnetic ordering of the $Nd^{3+}$ spins, as confirmed by the powder neutron diffraction and specific heat measurements discussed laterin this paper.

The inverse dc magnetic susceptibility data (shown in Fig. 2d) in the high temperature range $180 \leq T \leq 300$ K is fit with the Curie-Weiss law, $\chi(T) = C/(T - \theta_{CW})$ which gives a Curie constant, $C = 2.62(1)$ emu-K/mol Nd. The effective moment was calculated using $\mu_{eff}^2 = 3Ck_B/N_A$. The fit curve is shown by the solid line and gives $\theta_{CW} = -178(2)$ K and a reduced effective paramagnetic moment, $\mu_{eff} = 4.58(3)\mu_B$/Nd. We cannot determine the exact contribution of $Nd^{3+}$ and $Ru^{4+}$ from this measurement, however, assuming the magnetic moment of $Ru^{4+} \approx 1\mu_B$, as reported by Ito *et al.*, for $Nd_2Ru_2O_7$ pyrochlore [17], the contribution from $Nd^{3+}$ would be around $3.5\mu_B$. The estimated magnetic moment is very close to the expected paramagnetic value of ground state of $Nd^{3+}$ ion, $\mu_{eff} = g_J\sqrt{J(J+1)} = 3.62\mu_B$ (for $g_J = \frac{8}{11}$ and $J = \frac{9}{2}$). In antiferromagnetic systems, a reduced effective magnetic moment is usually observed due to zero-point quantum mechanical fluctuations worsened by frustration. Although there is normally a very small deviation from the expected effective moment value, as in $Gd_2Ti_2O_7$ [24], large deviations has also been reported in some systems including $Er_2Ti_2O_7$ [25]. To visualize magnetic frustration, the frustration index is given by $f = |\theta_{CW}|/T_N$ [26]. For moderately frustrated systems, $f \geq 3$ and in case of $Nd_2Ru_2O_7$ $f$ comes out to be 1.2 indicating that antiferromagnetic ordering is not strongly frustrated. One more thing we have noted in the fitting of the inverse susceptibility of $Nd_2Ru_2O_7$ is that the fitting curve deviates from the ideal Curie-Weiss law at lower temperatures, which is in good agreement with the earlier published reports [23], and indicates the presence of short-range ferromagnetic interactions at these temperatures.

As discussed in the heat capacity section below, we have estimated the first-excited crystal field level to be at about 188 K, therefore, because of the thermal population of higher CEF levels at $T \geq 100$ K, the estimated values of $\theta_{CW}$ and $\mu_{eff}$ for the Ising ground state obtained by the analysis of

$\chi(T)$ above are not accurate. To try to address this problem we fit $\chi(T)$ at temperatures below 20 K to a modified Curie-Weiss behaviour $\chi(T) = \chi_0 + C/(T - \theta_{CW})$, where the temperature independent term $\chi_0$ represents a Van Vleck contribution. We have fit $\chi^{-1}(T)$ in the temperature range $5 \leq T \leq 18$ K. The fit of $\chi^{-1}(T)$ by the modified Curie-Weiss law is shown by the solid line in the inset of Fig. 2(d) and yields C = 0.695(5) emu-K/mol Nd, $\theta_{CW}$ = +0.11(4) K and $\chi_0$ = 6.93(3) × 10$^{-3}$ emu/mol-Nd. The positive value of $\theta_{CW}$ indicates a weak ferromagnetic coupling between the Nd-moments. A positive value of $\theta_{CW}$ was also found for $Nd_2Hf_2O_7$ [20]. The value of C gives $\mu_{eff}$ = 2.37(4)$\mu_B$/Nd for Ising ground state of $Nd_2Ru_2O_7$.

Isothermal magnetization versus field curves for $Nd_2Ru_2O_7$ are shown in Fig. 3. At 2 K, the magnetization increases rapidly with increasing field up to 10 kOe and is a linear function of $H$. For $H \geq 10$ kOe, the magnetization becomes nonlinear as a function of $H$ and tends to saturate with a saturation magnetization of 1.1$\mu_B$/Nd at 50 kOe. The theoretical saturation magnetization for free $Nd^{3+}$ ions is $M_{sat} = g_J J\mu_B$ = 3.27$\mu_B$/Nd ($g_J$ = 8/11 and J = 9/2). Therefore, the observed saturation magnetization is about 34% of the free ion theoretical value. Such behaviour indicates a significant single-ion anisotropy, as expected for a local <111> Ising anisotropic system. A similar behaviour for $M$ is also observed in the case of $Nd_2Hf_2O_7$ [20]. With an increase in the temperature, the magnetization becomes a linear function of $H$ for a wider field range; however, the increase in $M$ with $H$ is not as steep.

We assume that the Kramers doublet ground state of $Nd^{3+}$ is described by an effective spin $S = ½$. The powder averaged and thermally averaged magnetization for an effective spin-half doublet ground state system with local <111> Ising anisotropy, assuming a transverse g-factor $g_\perp = 0$ and longitudinal g-factor $g_\parallel = g$ is given by expression [20, 27, 28]:

$$< M >/\mu_B = \frac{(k_B T)^2}{g\mu_B H^2 S} \int_0^{g\mu_B HS/k_B T} x\tanh(x)dx \qquad (1)$$

which can be integrated numerically, where $x = g\mu_B HS/k_B T$. In this expression, $H$ is a variable (in units of tesla) and $g$ is the only adjustable fitting parameter as all other terms are constants. The $M(H)$ data has been fit reasonably well by equation (1) as shown in Fig. 3, giving $g$ = 4.467(1) for $T = 2$ K, which is lower than the expected value for a pure $m_J = \pm\frac{9}{2}$ doublet, $g = 2g_J J$ = 6.54 ($g_J = \frac{8}{11}$, $J = \frac{9}{2}$), because of the mixing of the $m_J$ ground states by the crystal electric field. For $Nd_2Hf_2O_7$ $g$ is found to be 5.01 [20] and for $Nd_2Zr_2O_7$ it is found to be 4.793 [27]. For an effective factor $g$ = 4.467(1) and $S = ½$, the paramagnetic moment is expected to be $m = gS\mu_B$ = 2.23$\mu_B$/Nd. In the case of a powder sample, the effective magnetic moment is related to the g-factor and is given as $\mu_{eff} = \frac{\sqrt{3}}{2}\bar{g}\mu_B$, where $\bar{g} = (g_\parallel^2 + 2g_\perp^2)/3$. As $g_\perp = 0$ and $g_\parallel = 4.467(1)$, we obtain $\mu_{eff} = 2.23\mu_B/\text{Nd}$, which is close to the $\mu_{eff} = 2.37(4)\mu_B/\text{Nd}$ obtained above from the dc susceptibility analysis. Furthermore, it can be seen from the isothermal curve at $T = 2$ K in Fig. 3 that fitting with equation (1) predicts a saturation of moment at the high fields, however, the measured moment shows a little increase at the high field > 40 kOe. This behaviour of $M(H)$ suggests the presence of spin fluctuations or a non-Ising contribution. Similar behaviour for $M(H)$ was observed in the case of Pr$_2$Hf$_2$O$_7$ [29].

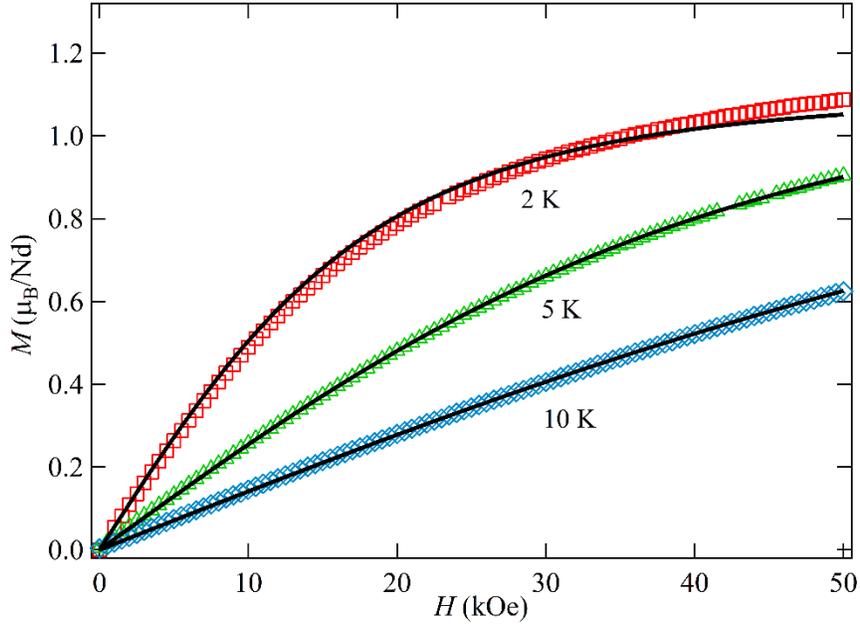

FIG. 3. Isothermal magnetization versus field curves of Nd$_2$Ru$_2$O$_7$ for a magnetic field $H \leq 50$ kOe measured at $T = 2$, 5, and 10 K. The solid curves represent fits of the $M(H)$ data using equation (1).

The nearest neighbour dipole-dipole interactions $D_{nn}$ can be estimated using our effective magnetic moment $\mu_{eff} = 2.37(4)\mu_B/\text{Nd}$ and lattice parameter $a = 10.3544(5)$ Å using equation [20,29]

$$D_{nn} = \frac{5}{3}\left(\frac{\mu_0}{4\pi}\right)\frac{\mu_{eff}^2}{r_{nn}^3} \approx 0.16 \text{ K}, \quad (2)$$

where $\mu_0$ is the magnetic permeability of the vacuum and $r_{nn} = \left(\frac{a}{4}\right)\sqrt{2}$ is the nearest neighbour distance. $D_{nn} = +0.12$ K has been reported for Nd$_2$Hf$_2$O$_7$ [20]. Following the procedure discussed in Refs. [20] and [30], we can roughly estimate the nearest neighbour exchange interaction between the <111> Ising moments by fitting the dc susceptibility data with the expression

$\chi(T) = \left(\frac{C_1}{T}\right)\left[1 + \frac{C_2}{T}\right]$, where $C_2$ can be decomposed as sum of dipolar and exchange interactions, $C_2 = \left(\frac{6S^2}{4}\right)[2.18D_{nn} + 2.67J_{nn}]$. By fitting the $\chi(T)$ data with this expression in the temperature range $5 \leq T \leq 18$ K, we get $C_2 =$ -0.57(2) K. Using the estimated values of $D_{nn} \approx 0.16$ K, we obtain $J_{nn} \approx$ -0.70 K. $J_{nn} \approx$ -0.77 K was observed for $Nd_2Hf_2O_7$ [20]. The value of $J_{nn}$ indicates that antiferromagnetic interactions dominate over dipolar interactions in $Nd_2Ru_2O_7$. The presence of antiferromagnetic exchange interactions in $Nd_2Ru_2O_7$ ultimately leads to a long-range ordered ground state of the $Nd^{3+}$ spins around 1.8 K as evidenced by the specific heat and neutron diffraction measurements. Hertog *et al.* [31] predicted by Monte Carlo simulations that long-range antiferromagnetic ordering with an all-in-all-out magnetic structure in pyrochores is possible for $J_{nn} / D_{nn} <$ -0.91. This theoretical condition is fulfilled for $Nd_2Ru_2O_7$ as $J_{nn} / D_{nn} \approx$ -4 < -0.91 and experimentally we have observed antiferromagnetic ordering of $Nd^{3+}$ spins with an all-in/all-out magnetic structure at 0.4 K as discussed in the neutron diffraction section.

### 3.3 Heat capacity measurements

Figure 4(a) shows the temperature variation of specific heat for $Nd_2Ru_2O_7$ in the temperature range 0.4 to 150 K in $H = 0$ Oe. The peak around 1.8 K represents the long-range ordering of $Nd^{3+}$ spins and transition near 144 K is due to the antiferromagnetic ordering of $Ru^{4+}$ ions, which is in agreement with the dc magnetic susceptibility measurements. An upturn has been observed in $C_p$ below 5 K (shown in the lower inset in Fig. 4(a)) indicating short-range magnetic correlations, consistent with the inverse susceptibility analysis. Previous studies [16] on the specific heat of $Nd_2Ru_2O_7$ have reported anomaly at 21 K, which is not present in our samples. This is possibly because the anomaly at 21 K is due to a canting of an already ordered moment and not due to a magnetic phase transition from a disordered to an ordered state. The temperature dependence of the specific heat above 10 K is well described by the relation $C_p = \beta T^3 + \delta T^5$; the electronic heat capacity coefficient $\gamma = 0$, because of the insulating nature of $Nd_2Ru_2O_7$ at low temperatures. A fit to $C_p/T$ versus $T^2$ plot with $C_p/T = \beta T^2 + \delta T^4$, as shown by the solid line in the inset of Fig. 4(a), for $9 \leq T \leq 22$ K yields $\beta = 5.6(4) \times 10^{-4}$ J/K$^4$-mol and $\delta = 3.7(5) \times 10^{-7}$ J/K$^6$-mol. The Debye temperature $\theta_D = 337(3)$ K is obtained from $\beta = (12/5)n\pi^4 R\theta_D^{-3}$, where n = 11 is the number of atoms per formula unit, $R = 8.314$ J/mol-K is the gas constant. The $\theta_D$ in pyrochlores has been found to be temperature dependent. In case of $Nd_2Hf_2O_7$, the low temperature $C_p(T)$ yields $\theta_D = 436$ K, while the high temperature $\theta_D = 785$ K [20]. For $Nd_2Zr_2O_7$, the low temperature $C_p(T)$ yields $\theta_D = 514$ K, while the high temperature $C_p(T)$ gives $\theta_D = 741$ K [21].

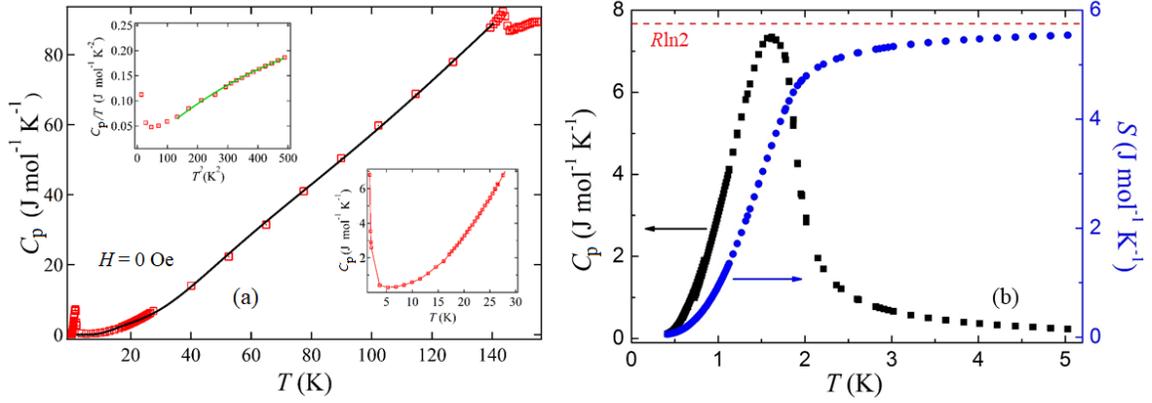

FIG. 4. (a) Temperature dependence of the specific heat of $Nd_2Ru_2O_7$ in the temperature range 0.4 to 150 K at $H = 0$ Oe. The solid line represents a fit to the Debye + Einstein + CEF model given by equation (3) for $2 \leq T \leq 140$ K. The upper inset shows $C_p/T$ versus $T^2$ for $10 \leq T \leq 22$ K. The solid line is a fit to $C_p = \beta T^3 + \delta T^4$ for $10 \leq T \leq 22$ K. The lower inset shows the upturn in $C_p$ observed before the $Nd^{3+}$ ordering. (b) Left axis: Temperature dependence of the specific heat of $Nd_2Ru_2O_7$ for $T < 5$ K. The peak at 1.8 K represents the long-range ordering of the $Nd^{3+}$ moments. Right axis: Temperature variation of the magnetic entropy of $Nd_2Ru_2O_7$. The increase in entropy saturates near 5 K and reaches the maximum value $R\ln 2$, which indicates the doublet ground state of the $Nd^{3+}$ ions. The raw $C_p$ data have been divided by 2 to obtain the specific heat per mole of Nd.

To obtain a more reliable value for $\theta_D$ we have analyzed the $C_p(T)$ data between 2 and 140 K using a combination of the Debye ($C_{V\,Debye}$) and Einstein ($C_{V\,Einstein}$) models of the lattice heat capacity, which take into account acoustic and optic phonon modes, respectively. In addition, we have considered the magnetic contribution $C_{CEF}$ to $C_p(T)$ of the $Nd^{3+}$ ions due to the crystal electric field of the $Ru^{4+}$ ions. Thus the $C_p(T)$ data are fit using the relation:

$$Cp(T) = mC_{V\,Debye}(T) + (1-m)C_{V\,Einstein}(T) + C_{CEF}(T). \quad (3)$$

The lattice contributions $C_{V\,Debye}$ and $C_{V\,Einstein}$ are given by [32]:

$$C_{V\,Debye}(T) = 9nR\left(\frac{T^3}{\theta_D^3}\right)\int_0^{\theta_D/T}\frac{x^4 e^x}{(e^x-1)^2}, \quad (4)$$

$$C_{V\,Einstein}(T) = 3nR\left(\frac{\theta_D^2}{T^2}\right)\frac{e^{\theta_E/T}}{(e^{\theta_E/T}-1)}, \quad (5)$$

where $\theta_E$ is the Einstein temperature. For a two level system the crystal field contribution, $C_{CEF}$ is given by [32]:

$$C_{CEF}(T) = R\left(\frac{\Delta^2}{T^2}\right)\frac{g_0\,g_1 e^{-\Delta/T}}{(g_0+g_1 e^{-\Delta/T})^2} \quad (6)$$

where $g_0$ is the degeneracy of ground state, $g_1$ is the degeneracy of first excited state and $\Delta$ is the energy gap between ground state and first excited state.

Inelastic neutron scattering (INS) measurements revealed [21] that the CEF splits the $(2J+1)$-fold degenerate ground state of $Nd^{3+}$ ($J = \frac{9}{2}$) into five doublets, therefore $g_0 = g_1 = 2$. The specific heat is fit with equation (3) as shown by the solid line in Fig 4(a) and we obtain $\Delta = 188(3)$ K. The lattice heat capacity is obtained from the difference $C_p(T) - C_{CEF}(T)$ and by fitting this difference using the Debye plus Einstein models in the temperature range $2 \leq T \leq 140$ K, we obtain $\theta_D = 790(6)$ K, $\theta_E = 102(5)$ K, with $m = 0.94$, indicating a 94% weight to the Debye term and a 6% weight to the Einstein term. The value of $\theta_D$ is comparable to other Nd-pyrochlores, where $\theta_D = 785$ K for $Nd_2Hf_2O_7$ [20] and $\theta_D = 741$ K for $Nd_2Zr_2O_7$ [21]. The value of $\Delta$ obtained from fitting is not accurate and can only be accurately obtained from INS measurements. In the case of $Nd_2Zr_2O_7$, $\Delta = 23.4$ meV $\approx 270$ K was obtained for $Nd^{3+}$ moments by INS measurements [21]. This indicates that there will be a thermal population of the higher CEF levels at $T > 100$ K for the $Nd^{3+}$ moments.

Fig. 4(b) shows the specific heat for $Nd_2Ru_2O_7$ (on the left-hand axis) in the low-temperature range ($T < 5$ K). The magnetic contribution ($C_m$) to the heat capacity has been obtained by subtracting lattice contribution [equivalent to $C_p(T)$ of non-magnetic $Y_2Ru_2O_7$] from the total heat capacity of $Nd_2Ru_2O_7$. The magnetic entropy is calculated by integrating $C_m(T)/T$ with respect to temperature using the formula: $\Delta S_m = \int_{T'}^{T''} \frac{C}{T} dT$, where $T'$ and $T''$ are the initial and final temperatures taken for the integration interval. As we can observe from the temperature dependence of entropy (shown on the right-hand axis in Fig. 4(b)), a zero-point or residual entropy of $\frac{R}{2}\ln\frac{3}{2}$ seen in spin ice does not appear in $Nd_2Ru_2O_7$ and $\Delta S_m$ saturates near 5 K attaining a value of 5.92 J/mol-K, which is close to the spin freezing condition $R\ln 2 = 5.76$ J/mol-K, and indicates $Nd^{3+}$ has a doublet ground state.

### 3.4 Neutron diffraction studies

We have studied the magnetic structure of $Nd_2Ru_2O_7$ at temperatures down to 0.4 K using powder neutron diffraction experiments. Neutron diffraction patterns recorded at 0.4 and 8 K are shown in Fig. 5(a) for 2θ range 19° to 133°. The data contain three peaks (denoted by *) from scattering due to Al and a peak at 83° (denoted by +) due to scattering from the Cu sample holder. The neutron diffraction data were refined using the *FULLPROF* suite [33]. Fig. 5(b) shows the results of the Rietveld refinement of the magnetic structure of $Nd_2Ru_2O_7$ at 0.4 K (after subtracting the 8 K nuclear pattern), where the red line is the experimentally observed data, the black line is the theoretically calculated intensity, and the line in blue represents the difference between the two patterns. We can observe the enhancement in intensity of the (220) and (113) magnetic peaks at

38° and 45° respectively (shown in the inset of Fig. 5(a)), which correspond to the most prominent peaks in the ordered state. A similar enhancement in the intensity of magnetic peaks is also observed for $Nd_2Hf_2O_7$ [20]. An additional magnetic peak at (420) has been observed near 63°, which confirms the antiferromagnetic ordering of the $Nd^{3+}$ spins. The magnetic peak (420) is not reproduced completely in the refinement, possibly because of its vicinity to an Al-peak. All the magnetic peaks correspond to the propagation vector $\mathbf{k} = (0, 0, 0)$. The propagation vector $\mathbf{k} = (0, 0, 0)$ was also found for the magnetic order in the case of $Nd_2Hf_2O_7$ [20]. In our synchrotron studies on $Nd_2Ru_2O_7$, down to 10 K, we observed a change in the lattice parameter, and a variation in the Ru-O, Nd-O and Ru-O-Ru bond lengths, which have been discussed in detail in our previous work on $Nd_2Ru_2O_7$ [14].

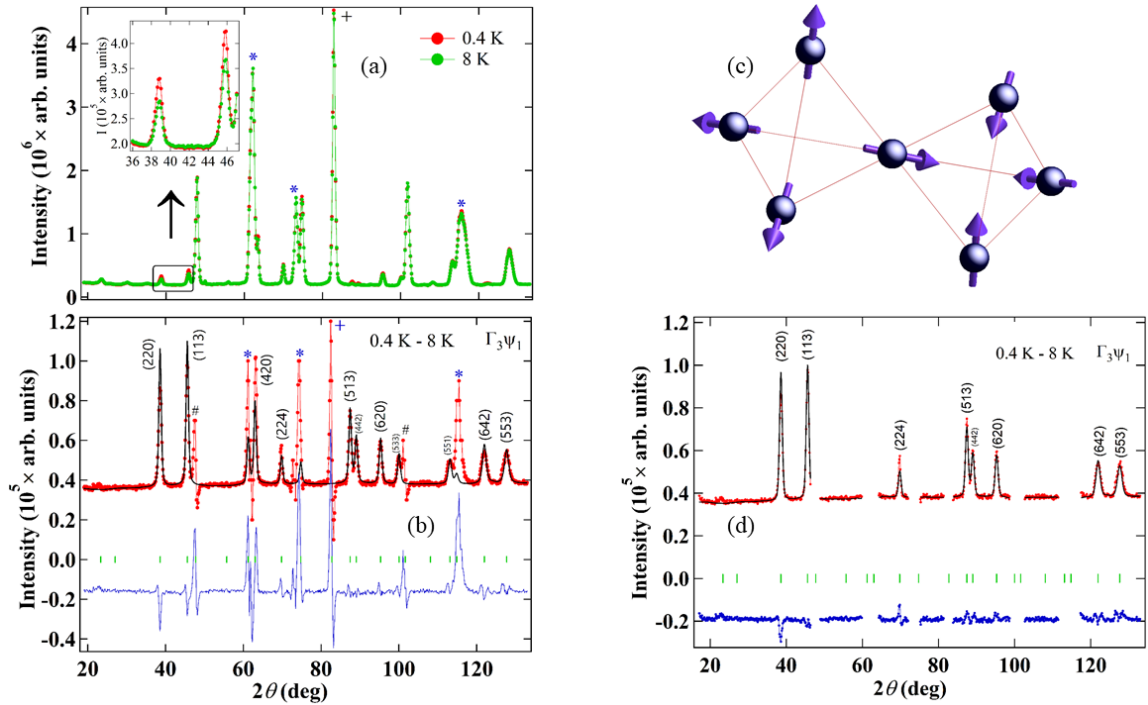

FIG. 5. (a) Neutron diffraction pattern of $Nd_2Ru_2O_7$ recorded at 0.4 and 8 K. (b) Rietveld refinement of the magnetic structure of $Nd_2Ru_2O_7$ at 0.4 K (after subtracting the 8 K nuclear pattern), where the observed data is shown in red, the black line is the theoretically calculated intensity, and the blue line shows the difference between the two. The green vertical lines indicate the positions of the Bragg peaks in the diffraction pattern. The neutron data are best modelled by the irreducible representation $\Gamma_3$ with an all-in-all-out spin configuration of the $Nd^{3+}$ ions. The peaks at $61^0$, $74^0$, and $115^0$ (denoted by *) are due to scattering from Al in the sample environment, the peak at $83^0$ (denoted by +) is due to the Cu sample holder. The peaks have been shifted upwards by 20000 for clarity. (c) The all-in all-out configuration of the magnetic moments. (d) Magnetic structure refinement profile of $Nd_2Ru_2O_7$ at 0.4 K (after subtracting the 8 K nuclear pattern), where the regions with background peaks have been excluded. The neutron wavelength used was 2.41 Å.

To further study the compatibility of the magnetic structure with the space group symmetry, we carried out a representational analysis using the SARA*h* software [34]. The symmetry analysis for the propagation vector $\mathbf{k} = (0, 0, 0)$ and space group $Fd$-$3m$ resulted in four nonzero irreducible representations (IRs) for the Nd(16$d$) site: $\Gamma_3$, $\Gamma_5$, $\Gamma_7$ and $\Gamma_9$ [35]. Table 2 shows basis vectors ($\psi$) corresponding to each irreducible representation. By combining the basis vectors of the IRs we can obtain different possible models of the magnetic structure. Out of four IRs, the magnetic structure of $Nd_2Ru_2O_7$ is best modelled by the irreducible representation $\Gamma_3$ (with a magnetic R-factor of 32.4%), with all-in-all-out spin configuration of the $Nd^{3+}$ ions as represented in Fig. 5(c). The ordered moment obtained from the refinement is 1.54(2)$\mu_B$/Nd. Fig. 5(d) shows the Rietveld refinement profile of the difference intensity 0.4 – 8 K (similar to Fig. 5b), where the regions with background peaks have been excluded. The magnetic R-factor is reduced to 10.7% and all the peaks due to magnetic structure have been reproduced e.g., (220), (113) etc. The ordered magnetic moment obtained from the refinement is reduced to 1.47(4)$\mu_B$/Nd.

In order to determine the magnetic structure from the presence or absence of the magnetic peaks at $2\theta < 50°$, we make the list of the non-zero magnetic peaks for each basis vector $\psi$, as shown in Table 3. We find that the magnetic structure corresponds to basis vector $\psi_1$. We can exclude the possibility of basis vectors $\psi_2$, $\psi_3$, $\psi_6$, $\psi_7$, $\psi_{12}$ and their allowable linear combinations, because corresponding magnetic structures lead to the sizable (111) magnetic peak, which is not observed in the experimental data.

For an effective factor $g = 4.467(1)$ and $S = ½$, the ordered moment is expected to be $m = gS\mu_B = 2.23\,\mu_B$/Nd. However, the effective magnetic moment of $Nd^{3+}$ at 0.4 K is estimated to be 1.54(2)$\mu_B$/Nd, which is smaller than the expected ordered moment. The reduced magnetic moment at the low temperatures indicates the presence of the quantum fluctuations which persist in the ordered state down to 0.4 K. A similar reduction in the magnetic moment of $Nd^{3+}$ is also observed in $Nd_2Hf_2O_7$ [20] and $Nd_2Zr_2O_7$ [21].

| IRs | $\psi$ | Nd1 | Nd2 | Nd3 | Nd4 |
|---|---|---|---|---|---|
| $\Gamma_3$ | $\Psi_1$ | (1 1 1) | (1 -1 -1) | (-1 1 -1) | (-1 -1 1) |
| $\Gamma_5$ | $\Psi_2$ | (1 1 -2) | (1 -1 2) | (-1 1 2) | (-1 -1 -2) |
|  | $\Psi_3$ | (1 -1 0) | (1 1 0) | (-1 -1 0) | (-1 1 0) |
| $\Gamma_7$ | $\Psi_4$ | (0 -2 2) | (0 2 -2) | (0 2 2) | (0 -2 -2) |
|  | $\Psi_5$ | (2 0 -2) | (-2 0 -2) | (-2 0 2) | (2 0 2) |
|  | $\Psi_6$ | (-2 2 0) | (2 2 0) | (-2 -2 0) | (2 -2 0) |
| $\Gamma_9$ | $\Psi_7$ | (2 0 0) | (2 0 0) | (2 0 0) | (2 0 0) |
|  | $\Psi_8$ | (0 1 1) | (0 -1 -1) | (0 -1 1) | (0 1 -1) |
|  | $\Psi_9$ | (0 2 0) | (0 2 0) | (0 2 0) | (0 2 0) |
|  | $\Psi_{10}$ | (1 0 1) | (-1 0 1) | (-1 0 -1) | (1 0 -1) |
|  | $\Psi_{11}$ | (0 0 2) | (0 0 2) | (0 0 2) | (0 0 2) |
|  | $\Psi_{12}$ | (1 1 0) | (-1 1 0) | (1 -1 0) | (-1 -1 0) |

Table 2: Irreducible representations (IRs) and associated basis vectors $\psi$ for the Nd (16d) sites in space group *Fd-3m* with propagation vector **k** = (0, 0, 0) for $Nd_2Ru_2O_7$ obtained from SARA*h* software[30]. The atoms of the non-primitive basis are defined according to Nd1: (0.5, 0.5, 0.5), Nd2: (0.5, 0.25, 0.25), Nd3: (0.25, 0.5, 0.25) and Nd4: (0.25, 0.25, 0.5).

| ψ | Magnetic peaks |
|---|---|
| $\Psi_1$ | (220), (113) |
| $\Psi_2$ | (111), (220), (113) |
| $\Psi_3$ | (111), (220), (113) |
| $\Psi_6$ | (111), (002), (220), (113) |
| $\Psi_7$ | (111), (220), (113), (222) |
| $\Psi_{12}$ | (111), (220), (113) |

Table 3: Magnetic peaks corresponding to each basis vector ψ.

### 3.5 Magnetic ordering in $Nd_2Ru_2O_7$

We have primarily focused on the low-temperature magnetic properties of $Nd_2Ru_2O_7$. In addition, we have tried to clarify the nature of the transitions at 21 and 146 K. Long-range magnetic order was observed at 1.8 K in all our measurements, indicating an antiferromagnetic ordering of the $Nd^{3+}$ moments below 1.8 K. Previously published reports have indicated that the transition at 146 K is due to an antiferromagnetic ordering of $Ru^{4+}$ 4$d$ spins, as a similar transition has been observed in $Y_2Ru_2O_7$, where there is no 4$f$ spin on the $A$-site [15, 23, 36]. The negative Curie-Weiss temperature, obtained from the fitting of inverse susceptibility curve also shows that the dominant magnetic interactions in $Nd_2Ru_2O_7$ are antiferromagnetic. However, contradictory reports are available in the literature regarding the ordering of $Ru^{4+}$ spins in $Nd_2Ru_2O_7$ and other analogous Ru-pyrochlores. Some studies have reported a glassy nature to the freezing of the $Ru^{4+}$ moments, which is evident from the reported bulk magnetic susceptibility measurements [15 - 17, 37, 38]. On the other hand, neutron diffraction experiments have shown long-range ordering of the $Ru^{4+}$ moments below the magnetic transition temperature of 146 K, with a long correlation length [15, 17].

Experimental evidence for spin-glass behaviour can be provided by ac susceptibility measurements, where the magnetic freezing temperature should vary with the frequency of the applied magnetic field [39]. In the case of $Nd_2Ru_2O_7$ it has been reported that the real, in phase component of the ac susceptibility, $\chi'_{ac}(T)$ shows that transition temperature at 146 K is independent of the frequency of the ac magnetic field, which is inconsistent with the behaviour expected for a spin-glass [23, 39]. Our results clarify that the peak around 146 K is due to the antiferromagnetic ordering of the $Ru^{4+}$ moments.

Contradictory conclusions are also drawn from the reports published in Refs. [16] and [23] regarding the origin of the cusp-like anomaly at 21 K and the divergence of the ZFCW-FCC curves below 21 K in $Nd_2Ru_2O_7$. Taira *et al.* [16] reported that the feature near 21 K might be due to a spin-glass transition, while Gaultois *et al.* [23] did not observe any transition around 21 K in their samples and concluded that the feature reported near 21 K in Ref. [16] resulted from the presence of a secondary phase in the samples. In the case of our samples, the presence of a secondary phase is excluded by an analysis of synchrotron x-ray data [14]. We have also performed ac susceptibility measurements in the temperature range 14 to 28 K as shown in the Fig. 6(a). The temperature dependence of the in-phase component of the ac susceptibility, $\chi'_{ac}(T)$ shows that transition temperature is independent of the frequency of the ac field. We have also performed powder neutron diffraction measurements (shown in Fig. 6(b)) below (18 K) and above (25 K) the transition at 21 K. We have normalized the intensity at 25 K with respect to the intensity at 18 K, and then calculated the subtracted intensity $I$ (18 K) – $I_{normalized}$ (25 K). We do not observe any noteworthy features in the subtracted intensity data. This indicates that there is no dramatic change in the magnetic structure of the $Ru^{4+}$ spins around 21 K. Furthermore, we have observed that the $Nd^{3+}$ spins remain paramagnetic down to 1.8 K, where magnetic susceptibility increases with decrease in the temperature. We therefore rule out the explanations proposed in Refs. [16] and [23] regarding the origin of the transition at 21 K. There is no possibility of any structural (or dramatic magnetic) transition at 21 K, as there is no clear feature in the heat capacity data around this temperature. Moreover, no additional diffraction peaks or phase transformation is observed in our x-ray diffraction experiments down to 10 K [14].

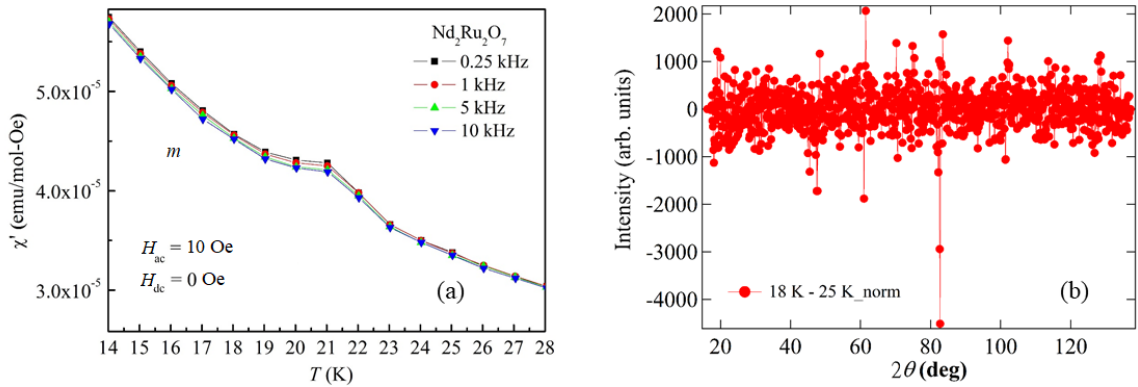

FIG. 6. (a): Temperature variation of ac susceptibility of $Nd_2Ru_2O_7$ at various frequencies with an ac field of 10 Oe. The position of the peak at 21 K does not change with the frequency of the ac field. $H_{dc}$ = 0 Oe. (b) The difference data of $I$(18 K) – $I_{normalized}$(25 K). No magnetic peaks are observed in the difference data indicating there is no discernable change in the magnetic structure around 21 K.

Previous studies on ruthenium pyrochlores have shown that the weak exchange coupling between the rare earth ($A^{3+}$) and $Ru^{4+}$ ions polarizes the $A^{3+}$ moment and consequently, the behaviour of $A^{3+}$ moments is not purely paramagnetic below the ordering of $Ru^{4+}$ spins [31, 36, 28]. Evidence for weak exchange coupling has been observed in our samples as discussed above in the dc susceptibility data analysis. We suspect a canting of $Ru^{4+}$ moments occurs because of the exchange coupling between the $Nd^{3+}$ and $Ru^{4+}$, which is evidenced by the positive $\theta_{CW}$ obtained in the dc susceptibility analysis at low temperatures. The presence of a weak ferromagnetic component associated with antiferromagnetic ordering of $Ru^{4+}$ spins is also evident from the sharply reduced Ru-O-Ru bond length at low temperatures in $Nd_2Ru_2O_7$, as explained in detail in our synchrotron studies [14]. Gaultois *et al.,* [23] also reported the presence of weak ferromagnetic interactions associated with the antiferromagnetic ordering of $Ru^{4+}$ spins below 146 K in $Nd_2Ru_2O_7$. In addition, Taira *et al.,* [16] have reported hysteresis curve at 5 K in $Nd_2Ru_2O_7$ samples, indicating the ferromagnetic component in the magnetic behaviour. The competition between the weak ferromagnetic and dominant antiferromagnetic interactions at the low temperatures might lead to a spin canting. This canting would then produce a splitting of the ZFCW and FCW curves at 21 K in the dc susceptibility. Thus, we suggest that the anomaly at 21 K is due to a weak ferromagnetic component produced by a polarization of the $Nd^{3+}$ moments and/or a canting of the ordered $Ru^{4+}$ moments. In principle, if some canting is present, the powder neutron diffraction patterns should not be the same at 18 and 25 K. However, the change in Ru-O-Ru bond angle is very small (≈ 1 degree) as observed in our synchrotron studies [14]. Therefore, the ferromagnetic component is very weak which will produce weak magnetic signal that might not be detected in the powder neutron diffraction experiments. That is why there is no noticeable difference between neutron spectra at 18 ad istead either said that there is a competition nd 25 K.

The weak exchange coupling between the rare earth $A^{3+}$ and the $Ru^{4+}$ ions below the ordering temperature of the $Ru^{4+}$ moments [40], eventually leads to the antiferromagnetic ordering of $Nd^{3+}$ spins below 1.8 K. This exchange dominates any dipole-dipole interactions between the $Nd^{3+}$ ions as discussed in the dc susceptibility analysis above. This result is supported by neutron diffraction experiments on the related systems $Er_2Ru_2O_7$ [40, 41] and $Tb_2Ru_2O_7$ [42], which confirm the long-range ordering of $A^{3+}$ *4f* spins. Ordering of $Tb^{3+}$ lattice is observed in the ruthenate pyrochlore $Tb_2Ru_2O_7$ at 3.5 K, however, no such ordering of the $Tb^{3+}$ sublattice is observed in the corresponding titanate pyrochlore $Tb_2Ti_2O_7$ down to 17 mK. Therefore, it is assumed that ordering of $Tb^{3+}$ moments in $Tb_2Ru_2O_7$ happens because of the internal field of $Ru^{4+}$ ions, which is supposed to relieve the frustration. A similar conclusion was arrived at for $Ho_2Ru_2O_7$ [43].

## 4. Summary

Polycrystalline pyrochlore $Nd_2Ru_2O_7$ has been prepared and examined using a combination of x-ray and neutron diffraction, magnetic, and heat capacity studies. A combination of ac and dc magnetic susceptibility measurements reveal three magnetic transitions in $Nd_2Ru_2O_7$ at 146, 21 and 1.8 K that are associated with an antiferromagnetic ordering of the $Ru^{4+}$ moments, a weak ferromagnetic signal attributed to a canting of the $Ru^{4+}$ ions and perhaps a polarization of the $Nd^{3+}$ moments, and a long-range ordering of the $Nd^{3+}$ moments, respectively. $M(H)$ data indicate the Nd moments exhibit a local <111> Ising anisotropic behaviour with an effective $g = 4.467(1)$ at $T = 2$ K, and an effective magnetic moment $\mu_{eff} = 2.33\mu_B/Nd$. The low-temperature dc susceptibility yields an effective moment $\mu_{eff} = 2.37(4)\mu_B/Nd$ for the Ising ground state and the positive value of $\theta_{CW}$ indicates the presence of weak ferromagnetic interactions between the Nd moments, although these moments eventually order antiferromagnetically below 1.8 K. The fitting of the $C_p(T)$ data using a Debye + Einstein + CEF model predicts the separation between ground state doublet and first excited state to be around 188(3) K. The magnetic entropy released up to 5 K is $R\ln2$, suggesting the $Nd^{3+}$ has a doublet ground state.

Above 9 K the specific heat data are well described by the relation $C_p = \beta T^3 + \delta T^5$, and a fit between 9 and 22 K yields $\beta = 5.6(4) \times 10^{-4}$ J/K$^4$-mol giving a Debye temperature $\theta_D = 337(3)$ K. A more accurate value for $\theta_D = 790(6)$ K is obtained by analysing the $C_p(T)$ data using a combination of a Debye and an Einstein model of the lattice heat capacity, and an additional magnetic contribution from the $Nd^{3+}$ ions due to the crystal electric field of $Ru^{4+}$ ions. This is comparable to the $\theta_D$ values for similar pyrochlores.

Powder neutron diffraction and our previous synchrotron x-ray diffraction measurements reveal the nature of the magnetic ordering transitions and the presence of lattice distortions accompanying the magnetic transitions. The peaks at 47° and 101° (denoted by #) in the powder neutron pattern might be present due to the distortion of the crystal structure at low temperatures. An additional magnetic peak at (420) has been observed at 63°, which confirms the antiferromagnetic ordering of $Nd^{3+}$ spins. The neutron diffraction data of $Nd_2Ru_2O_7$ is best modelled by the irreducible representation $\Gamma_3$. The magnetic moment of $Nd^{3+}$ ion at 0.4 K is estimated as $1.54(2)\mu_B$ and the magnetic structure is all-in all-out as determined by neutron experiments. The reduced magnetic moment indicates the strong quantum fluctuations, which persist down to 0.4 K. The reduction in the magnetic moment and the presence of quantum fluctuations might be because of the dipolar-octupolar nature of Kramers doublet ground state of $Nd^{3+}$ [22]. Due to the presence of the octupolar term, the $Nd_2Ru_2O_7$ may not behave strictly like a dipolar system and this non-Ising contribution can cause quantum fluctuations. Further theoretical and experimental work need to done to confirm this.

**Acknowledgments**

We thank the Ministry of Science and Technology, Taiwan who supported the work, via the grants: MOST 104-2221-M-006-010-MY3, MOST 104-2119-M-006-017-MY3. We thank J. S. Gardner for fruitful discussions.